\documentclass[10pt]{wlscirep}

\title{Non-criticality of interaction network over system's crises: A percolation analysis}
\author[1, *]{Amir Hossein Shirazi}
\author[2, 3, 4, +]{Abbas Ali Saberi}
\author[1, 4]{Ali Hosseiny}
\author[1]{Ehsan Amirzadeh}
\author[1]{Pourya Toranj Simin}
\affil[1]{Department of Physics, Shahid Beheshti University, G.C., Evin, Tehran 19839, Iran}
\affil[2]{Department of Physics, University of Tehran, Tehran 14395-547, Iran}
\affil[3]{Institut f\"ur Theoretische Physik, Universitat zu K\"oln, Z\"ulpicher Strasse 77, 50937 K\"oln,Germany}
\affil[4]{School of Physics and Accelerators, Institute for research in Fundamental Science (IPM) PO Box 19395-5531, Tehran, Iran}

\affil[*]{amir.h.shirazi@gmail.com}
\affil[+]{ab.saberi@ut.ac.ir}

 \begin{abstract}
Extraction of interaction networks from multi-variate time-series is one of the topics of broad interest in complex systems. Although this method has a wide range of applications, most of the previous analyses have focused on the pairwise relations. Here we establish the potential of such a method to elicit aggregated behavior of the system by making a connection with the concepts from percolation theory. We study the dynamical interaction networks of a financial market extracted from the correlation network of indices, and build a weighted network. In correspondence with the percolation model, we find that away from financial crises the interaction network behaves like a critical random network of Erd\H{o}s-R\'{e}nyi, while close to  a financial crisis, our model deviates from the critical random network and behaves differently at different size scales. We perform further analysis to clarify that our observation is not a simple consequence of the growth in correlations over the crises.
\end{abstract}

\date{\today}

\begin{document}
\maketitle

\section{Introduction}

Classical inverse statistical mechanics has been applied to determine properties of pairwise interaction potentials from ensemble properties in complex networks \cite{lezon}. The maximum entropy model characterizes the correlation
structure of a network activity without assumptions
about its mechanistic origin and enables predictions of the collective effects \cite{bialek}. These maximum entropy models are equivalent
to Ising models in statistical physics and, as we show in the following, it allows us to explore critical behavior of the network.

Despite wide range application of maximum entropy model in the recent decade \cite{lezon, bialek, roudi, jafari, roudiEconomy}, the main focus has been made on characterization of specific interaction between two elements or community structure of networks. However, global dynamical response of an evolving network based on its inherent symmetries has not been addressed yet. Here we apply percolation theory to measure the strength of interactions in a time-dependent evolving network of financial market and show that away from financial crisis the interaction network is self-similar and exhibits a geometrical criticality at a certain size-independent interaction threshold while during the crisis the network responses differently at different size-scales. 

Percolation theory \cite{saberiPercolationRev} is the simplest fundamental model in statistical physics that displays phase transitions and explains the behavior of connected clusters in a random graph. The geometric critical behavior is dominated by the emergence of a giant connected component which controls the global response of the network. Resilience of networks under attacks \cite{barabasi_attack, havlin_attack}, spreading phenomena and epidemics \cite{newman_book, barabasi_book, vespignani_epidemy, vespignani_review} are examples of diverse problems that can be treated using percolation theory.  

The concept of random graphs \cite{newman_book, barabasi_book} was put forward by Erd\H{o}s-R\'{e}nyi \cite{erdos} who introduced the simplest imaginable random graph ER$_n$($p$) including $n$ vertices in which an edge is placed between any pair of distinct vertices with some fixed probability $p$. This random graph exhibits a continuous phase transition at a critical threshold $p_c$ which leads to a sharp global connectivity of the network components and serves as the mean-field model of percolation. 

We find that despite seemingly strong correlations in the interaction network of financial market away from a global crisis, it can be modeled, to a good extent, by an ER$_n$($p$) random graph of $n$ interacting stock markets when the control parameter $p$ is taken to be the strength of pairwise interactions.  

In this paper, we study the critical behavior of a financial network consist of about 400 indices (or vertices) in S$\&$P 500 whose activities are available as time series. We first build up a time-dependent correlation matrix of the stocks from the time-series. Then, we extract the interaction matrix among system’s elements in which the non-direct correlations are eliminated. This interaction matrix is the adjacency matrix of elements’ interaction network. Given the interaction network of a system, we can reduce the full system to some disjoint components which have positive intra-interactions (compared with a given strength threshold, see below). 
The collective and large-scale information of the system is somehow encoded in the statistics of the largest component which can also control the dynamics of the system. This collective behavior can also result in large-scale deviations in system states. For example, in financial market all stock indices may fall down and influence global index of the market \cite{sornette2}. 

Our analysis of the time series for indices in S$\&$P 500 as the system elements and its mapping to the percolation problem on networks, unravels that for a network of stock markets, the dynamics can be well modeled by a critical random network theory of ER$_n$($p$) away from a global financial crisis, while around and at the crisis the network departures from criticality. This observation is in contrast with the ordinary critical phenomena in which the large-scale fluctuations play a crucial role in the behavior of systems in the vicinity of the critical state and the fluctuations are actually responsible for the criticality. Despite large fluctuations in stock’s prices over a crisis period (Fig. 1, light blue bars), the underlying network model shows a non-critical behavior and fluctuations drive the system out of criticality. 

\section{Data preparation}	

\begin{figure}[ht]
\centering
\includegraphics[scale = 0.5]{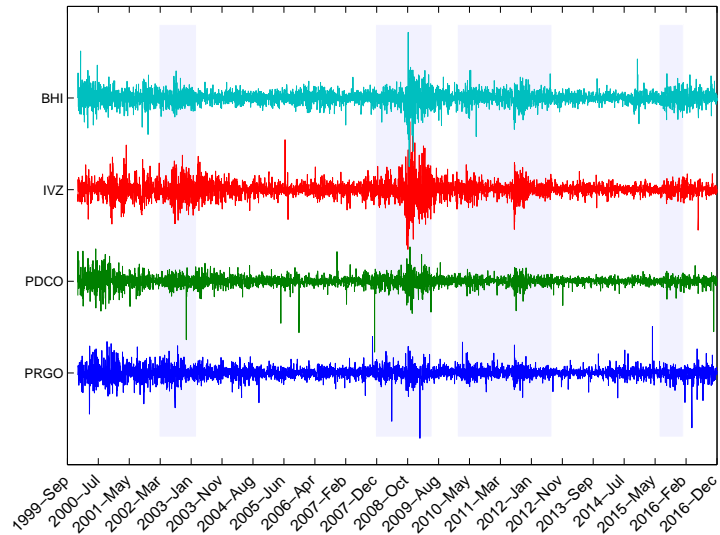}
\caption{Log-return of prices for $4$ sample stocks are shown in different colors. The columnar shadow windows show the periods of major financial crises.}
\label{fig:sample}
\end{figure}

We analyse the available data for "adjusted closing prices" in S\&P 500 index between 2000 and 2017. The time-series $\overrightarrow{X}(t) = (X_1 (t), X_2(t), ..., X_n(t))^T$ of $n=396$ stocks' prices are extracted from \textit{finance.yahoo.com}. We only consider the data for working days and use linear interpolation method to treat the sparsity of our data. For each stock $i$, as shown in Fig. \ref{fig:sample}, we work with the normalized log-return $x_i(t)$ of data \cite{mantegna_book}  defined as 
\begin{equation}
x_i(t)=(x'_i(t)-\mu(t))/\sigma(t) \quad with \quad x'_i(t) = \log X_i(t)-\log X_i(t-1) ,
\end{equation}
where $\mu(t)$ and $\sigma(t)$ denote the average and standard deviation of $x'$, respectively.

\section{Interaction network}

For a given time $t$ and a time window $\tau$ (see Fig. \ref{fig:sample}), we construct a multivariate $n\times \tau$ matrix $D_\tau(t)$ for $n$ time series as
\begin{equation}
D_\tau(t) = \left[\begin{array}{cccc}
x_1(t-\tau+1) & x_1(t-\tau+2) & \cdots & x_1(t) \\
x_2(t-\tau+1) & x_2(t-\tau+2) & \cdots & x_2(t) \\
 & \vdots & & \\
x_n(t-\tau+1) & x_n(t-\tau+2) & \cdots & x_n(t) \\
\end{array}
\right].
\end{equation}

We then build up the correlation matrix $C_\tau(t) $ among the time-series for each time window as
\begin{equation}\label{eq.corr}
C_\tau(t) = D_\tau(t) \cdot D_\tau(t)^T. 
\end{equation}
In order to also monitor the time evolution of the correlation matrix we move the time $t$ by steps of duration $30$ days over the whole period between 2000 to 2017. 

Based on the above correlation matrix Eq. (\ref{eq.corr}), we extract the interaction matrix $J_\tau(t)$ \cite{bialek, lezon, roudi, jafari, roudiEconomy} 
\begin{equation}
J_\tau(t) = \left[\begin{array}{cccc}
j_{11} & j_{12} & \cdots & j_{1n} \\
j_{21} & j_{22} & \cdots & j_{2n} \\
\vdots & \vdots & \ddots & \vdots \\
j_{n1} & j_{n2} & \cdots & j_{nn} \\
\end{array}
\right],
\end{equation}
whose symmetric elements $j_{lk}=j_{kl}$ represent the strength of interaction between stocks $l$ and $k$ in the time window [t,t+$\tau$).  Due to the finite sample size in our data sets, we used "Graphical LASSO" technique \cite{tibshirani, sander} to evaluate interaction matrix $J_\tau(t)$ (we set regularization penalty of GLASSO to 0.1. We have also used the output $\Theta$ of graphical LASSO, as an estimator of inverse co-variance matrix. This matrix appears in multi-variate Gaussian distribution and determines the strength of the interactions among the different dimensions of PDF.  In comparison with the Ising model, the coupling coefficients are minus sign of this matrix \cite{lezon, tibshirani, sander}.). 

Interaction matrix has an important advantage over the correlation matrix in which all mediated correlations are eliminated in the interaction matrix. Therefore, if a high correlation between indices A and B would be due to their high correlation with an index C, this effect will be eliminated in the interaction matrix.

This concept is related to the partial correlation matrix in multivariate Gaussian noise \cite{tibshirani}. It is also related to the maximum entropy network in the inverse statistical physics \cite{lezon}. Based on this fact, elements of the interaction matrix are independent of each other and one can remove them during the attack process (percolation model).

\section{Percolation model analysis}

For each interaction matrix $J_\tau(t)$ at time $t$, we have an adjacent network $G(t) = (n, E)$ with weighted links. In order to establish a connection with ordinary percolation problem on networks, we consider a threshold $\theta$ for the weights on the links to transform the network to an unweighted one $G_{\theta}(t)=(n, E_{\theta})$. This means that we remove the links whose weights are smaller than the considered threshold $\theta$, and keep other links in the network by setting their weights equal to $1$ \cite{saberi1, saberi2}, i.e.,
\begin{equation}
\text{for each}\;e_{ij} \in E_\theta,\; \; \; e_{ij} = 
\left\{\begin{array}{rl}
	0 & j_{ij}<\theta \\
	1 & j_{ij}\geq\theta.	
\end{array}\right .
\end{equation}
 
\begin{figure}
\centering
\includegraphics[scale=1.5]{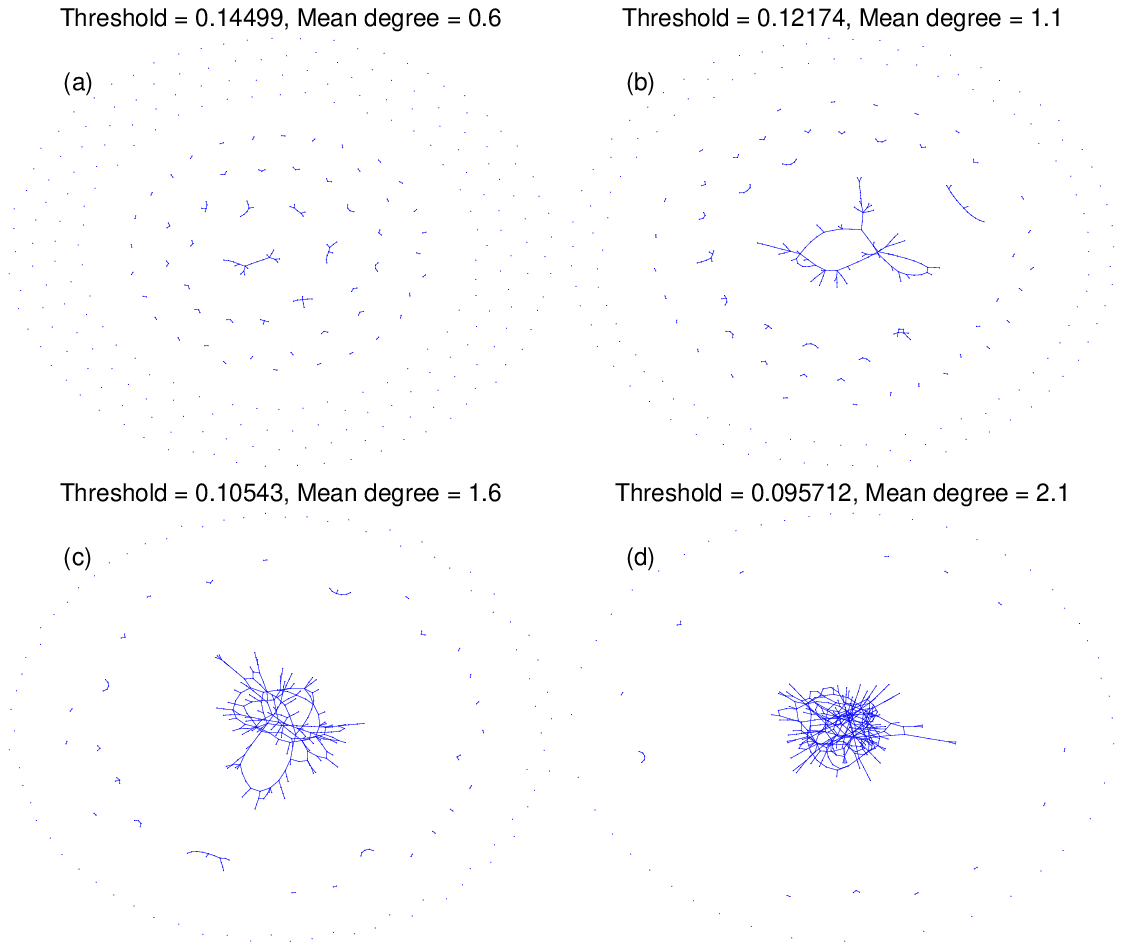}

\caption{A sample interaction network with four different threshold levels on links' weight. The birth of the giant connected component can be seen by decreasing the threshold level.}
\label{fig:networkVisual}
\end{figure}

Figure \ref{fig:networkVisual} illustrates this procedure on a sample network for four different threshold levels. It demonstrates how the giant component is born as the threshold level decreases. For a given graph $G_\theta$ we define the percolation strength $P_{\infty}(\theta)$ as the probability that a randomly chosen node belongs to the largest connected component of the graph. Based on the percolation theory, the network may undergo a geometrical phase transition during which the size of the giant connected component jumps to a size of order $\mathcal{O}(n)$ \cite{newman_book, barabasi_book}. To be more comparable with the ER$_n$ random networks, we interchange our percolation parameter $\theta$ with the corresponding mean degree $\bar{k}(G_\theta)$ of the graph. For the ER$_n$ graphs, the critical point is known to be $\bar{k}_c=1$ \cite{newman_book, barabasi_book}.

\begin{figure}
\centering
\includegraphics[scale=1.5]{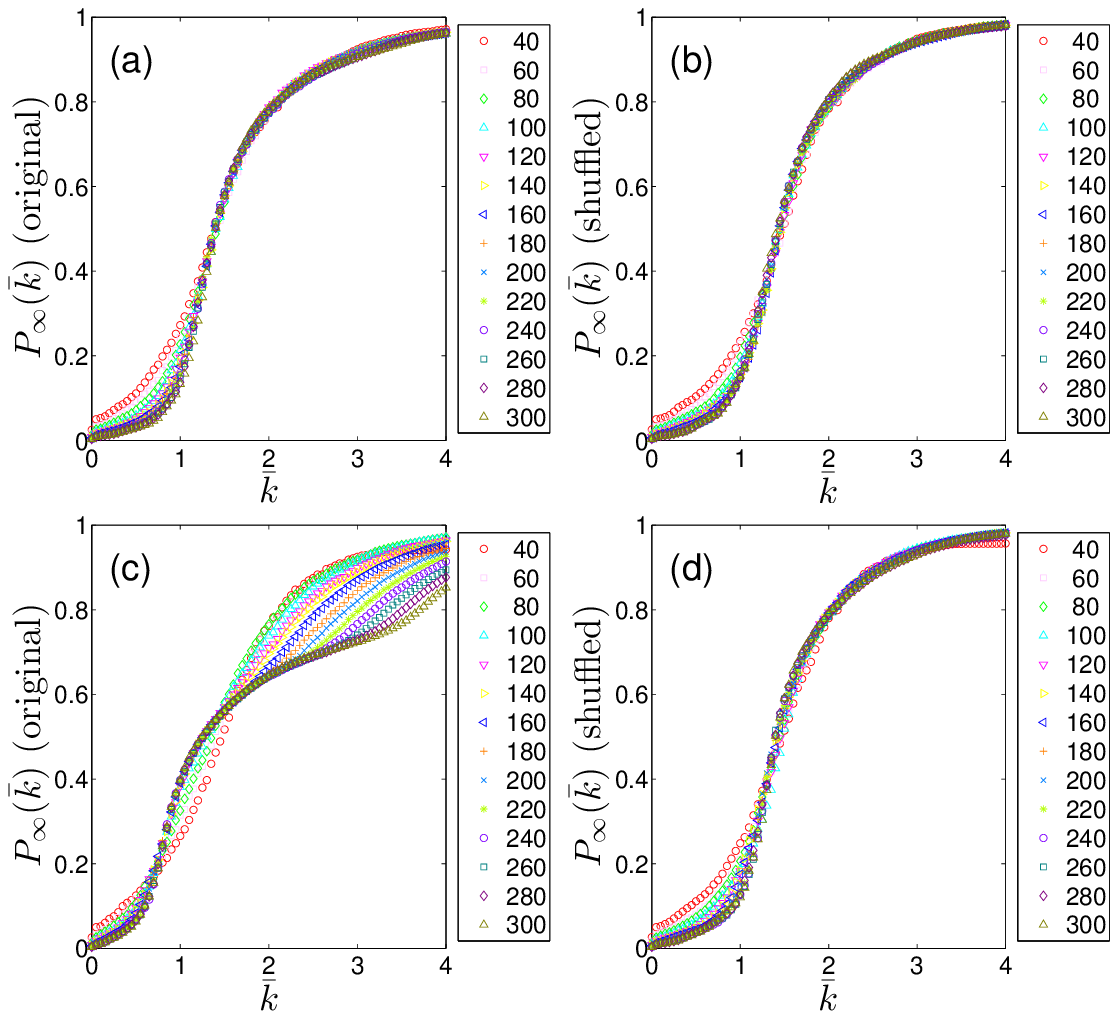}

\caption{The giant component probability $P_{\infty}(\bar{k})$, as a function of the mean degree $\bar{k}$. (a) An interaction network far from crises and (b) its shuffled network (2000 Jan-May). (c) A network close to a crisis and (d) its shuffled network (2008 Jul-Nov).}
\label{fig:GC}
\end{figure}

The first quantity of interest is $P_{\infty}(\bar{k})$ for a given interaction network $G(t)$.
In order to investigate the self-similarity of the network at different length scales, we also consider several sub-graphs $g(t, s)$ of size $s\le n$ which are randomly chosen from the original network $G(t)$. We then measure $P_{\infty}(\bar{k})$ for different size scales which is averaged over an appropriate number (about 100) of independent realizations for each size $s$. In Fig. \ref{fig:GC}, we present the results of our measurements of $P_{\infty}(\bar{k})$ for two time periods far from (Fig. \ref{fig:GC}-a) and close to (Fig. \ref{fig:GC}-c) the financial crisis (In the Supplementary we present the same quantity measured for various time periods). As it is obvious from the Figs. \ref{fig:GC}-a and \ref{fig:GC}-c, $P_{\infty}(\bar{k})$ clearly behaves differently within these two time periods.  Our further investigation shows that the difference between these two relies on the difference in the structure of the interaction network. To this aim, we shuffle the networks and repeat our analysis. Since a link's weight comes from the correlation of agents actions on two stocks, shuffling links is equivalent to a situation where agents buy or sell randomly. We find that away from the crises, the shuffled network is very similar to the original network (Fig. \ref{fig:GC}-b) as for the ER$_n$ random networks, while close to the crises, the original network substantially deviates from the shuffled one (Fig. \ref{fig:GC}-d).

\begin{figure}
\centering
\includegraphics[scale=1.5]{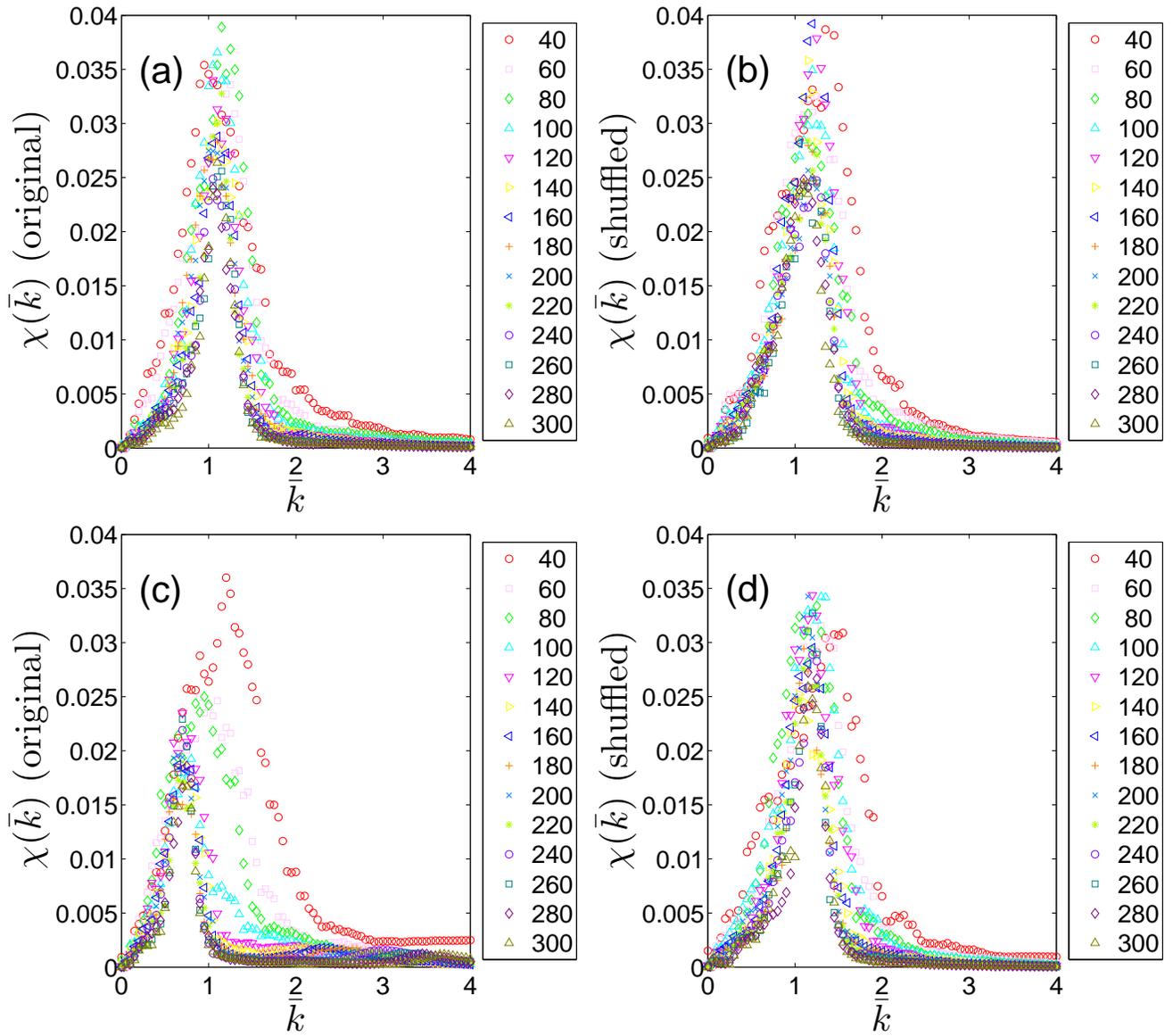}

\caption{The mean cluster size $\chi(\bar{k})$ as a function of the mean degree $\bar{k}$. (a) An interaction network far from crises and (b) its shuffled network (2000 Jan-May). (c) A network close to a crisis and (d) its shuffled network (2008 Jul-Nov).}
\label{fig:chi}
\end{figure}

The other quantity of interest is the mean cluster size (or susceptibility) $\chi(\bar{k})$ defined as \cite{castellano}
\begin{equation}
\chi(\bar{k}) = \frac{\langle P^i_{\infty}(\bar{k})^2 \rangle}{\langle P^i_{\infty}(\bar{k}) \rangle},
\end{equation}
where $\langle\dots\rangle_i$ denotes averaging over the number (about 100 for each size $s<n$) of independent realizations and $P^i_{\infty}$ is the percolation strength computed for the $i$-th realization. 
Figure \ref{fig:chi} shows the results of our computations of $\chi(\bar{k})$ for two time periods, as in the Fig. \ref{fig:GC} above, far from (Fig. \ref{fig:chi}-a) and close to (Fig. \ref{fig:chi}-c) the financial crisis (Supplementary presents the same quantity for various time periods). Figure \ref{fig:chi}-a indicates that far from the crisis, all curves $\chi(\bar{k})$ for different size scales maximize near a single size-independent critical point $\bar{k}_c\approx 1$, which is very similar to its shuffled variant shown in the Fig. \ref{fig:chi}-b. Close to the crisis, in contrast, $\chi(\bar{k})$ behaves differently at different size scales with an observable shift from the critical point (Fig. \ref{fig:chi}-c). It also differs from its shuffled version as is evident in the (Fig. \ref{fig:chi}-d).

To quantify the amount of deviation from the critical behavior near the crisis, let us now measure the difference between the strength of the giant component and its prediction based on the random network theory \cite{newman_book, barabasi_book} which states that it is possible to compute the giant component probability $P^{theo}_\infty(\bar{k})$ by solving the following self-consistent equation (see the Supplementary for more description)
\begin{equation}
P^{theo.}_\infty(\bar{k}) = 1 - e^{-\bar{k}P^{theo.}_\infty(\bar{k})}
\end{equation}
Next, we measure the mean absolute difference $d(t, s)$ between the theory and our numerical computations as
\begin{equation}\label{Eq:diff}
d(t, s) = \langle \left| \langle P_\infty(\bar{k})\rangle_i-P^{theo.}_\infty(\bar{k}) \right| \rangle_{\bar{k}}.
\end{equation}

We have plotted $d(t, s)$ for different sub-graph sizes $s$ in Fig. \ref{fig:compareER} for both original and shuffled networks. 
Close to the crisis periods (highlited in the Fig. \ref{fig:compareER} by light blue bars)  $d(t, s)$ increases. It shows that the critical behavior of the interacting financial network (which shown to behave like a random network) disappears close to a crisis.





\section{Conclusion}

Efficiency of financial markets is a hot debate in financial economics. Some studies support the hypothesis of market efficiency---see for example \cite{fama65,fama70}. According to this hypothesis, the stock prices fluctuate mostly like uncorrelated random variables. In other words, it states that it is impossible to extract information from the past history of the prices of a stock to predict its future and earn money. 

Despite early observations which supported the hypothesis of efficiency, some later works challenged it and revealed deviations from it. Although it was hard to find correlations in time series of an individual stock, noticeable information could be captured if some other parameters such as cross correlations or earning price ratio were brought into account---see for example \cite{lo,Jegadeesh}, \cite{mantegna_book} and references therein.

\begin{figure}
\centering
\includegraphics[scale=0.7]{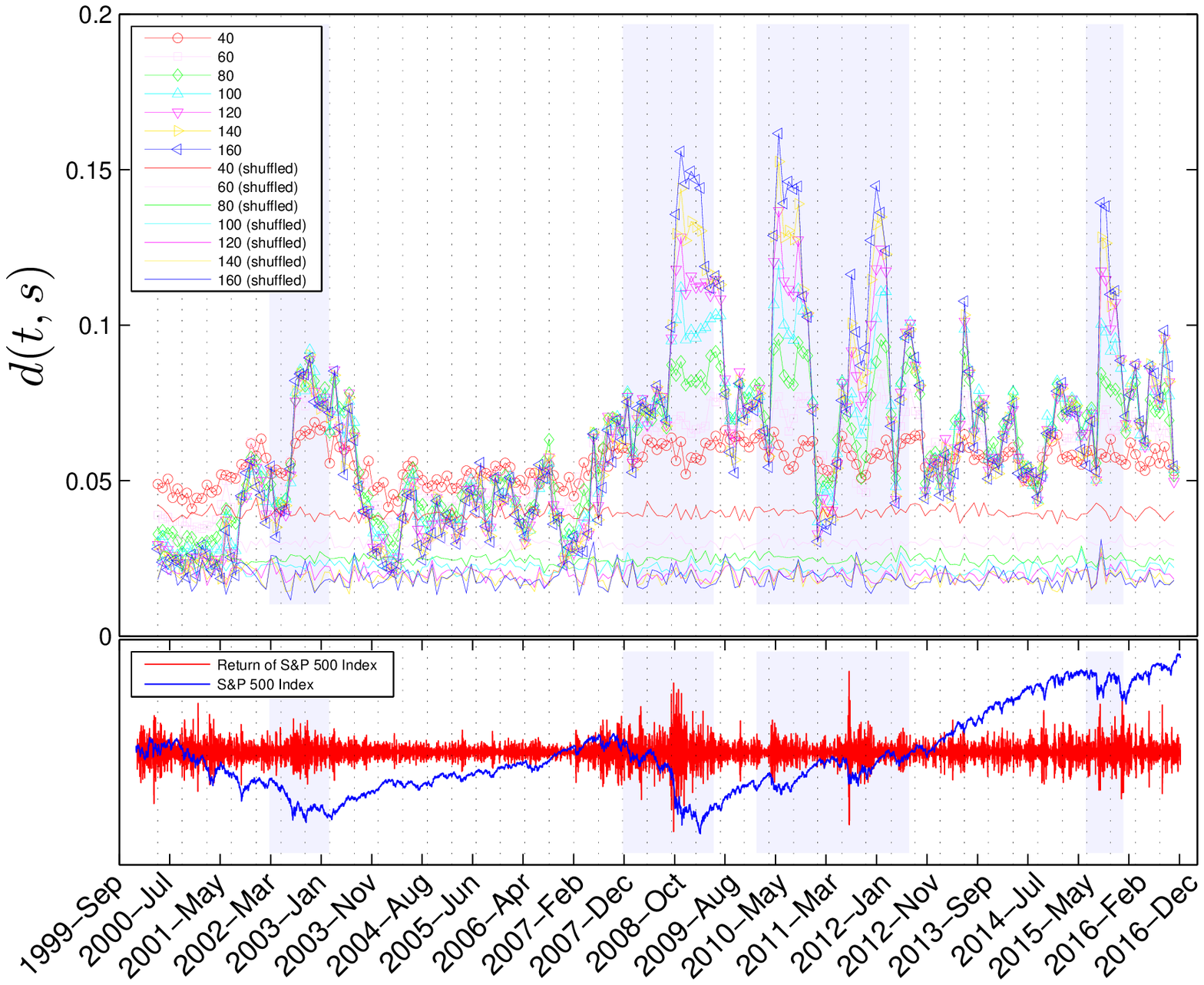}
\caption{The mean absolute difference between theory and our computations for the giant component probabilities (top panel) for time period $\tau=90$ and different sub-graph size (different symbols of different colors). Larger sub-graph sizes exhibit larger fluctuations in time. The thin solid curves in different colors show the same quantity for the corresponding shuffled network of each size which are comparatively less fluctuating around a mean value. The bottom panel shows the data for S\&P500 index (blue curve) and its increments (red curve). The vertical bars in light blue show four different major crisis periods: (i) Stock market downturn of 2002, (ii) Financial crisis of 2007-08, (iii) 2010 Flash Crash and August 2011 stock markets fall, and (iv) 2015-16 stock market sell off.}

\label{fig:compareER}
\end{figure}

The studied methods have mostly focused on the individual indices or their cross correlations and the analysis based on an aggregated behavior is still lacking. In the present work, we studied the global behavior of indices as an example of an interaction network by using the concepts of percolation theory. We find that away from financial crises the interaction network behaves like a random network of Erd\H{o}s and R\'{e}nyi \cite{erdos} which similarly exhibits the properties of scale invariance and self-similarity at the critical point of a continuous phase transition. When the financial market approaches a crisis, our observation is that the interaction network model deviates from the critical random network and looses its scale invariance, i.e., the system behaves differently at different size scales. The deviations are summarized in Fig. \ref{fig:compareER} in which our data signals at major crashes of the markets namely, "Stock market downturn of 2002", "Financial crisis of 2007-08", "2010 Flash Crash", "August 2011 stock markets fall", and "2015-16 stock market sell off", by a noticeable growth of difference with respect to the random networks. 

\begin{figure}
\centering
\includegraphics[scale=0.7]{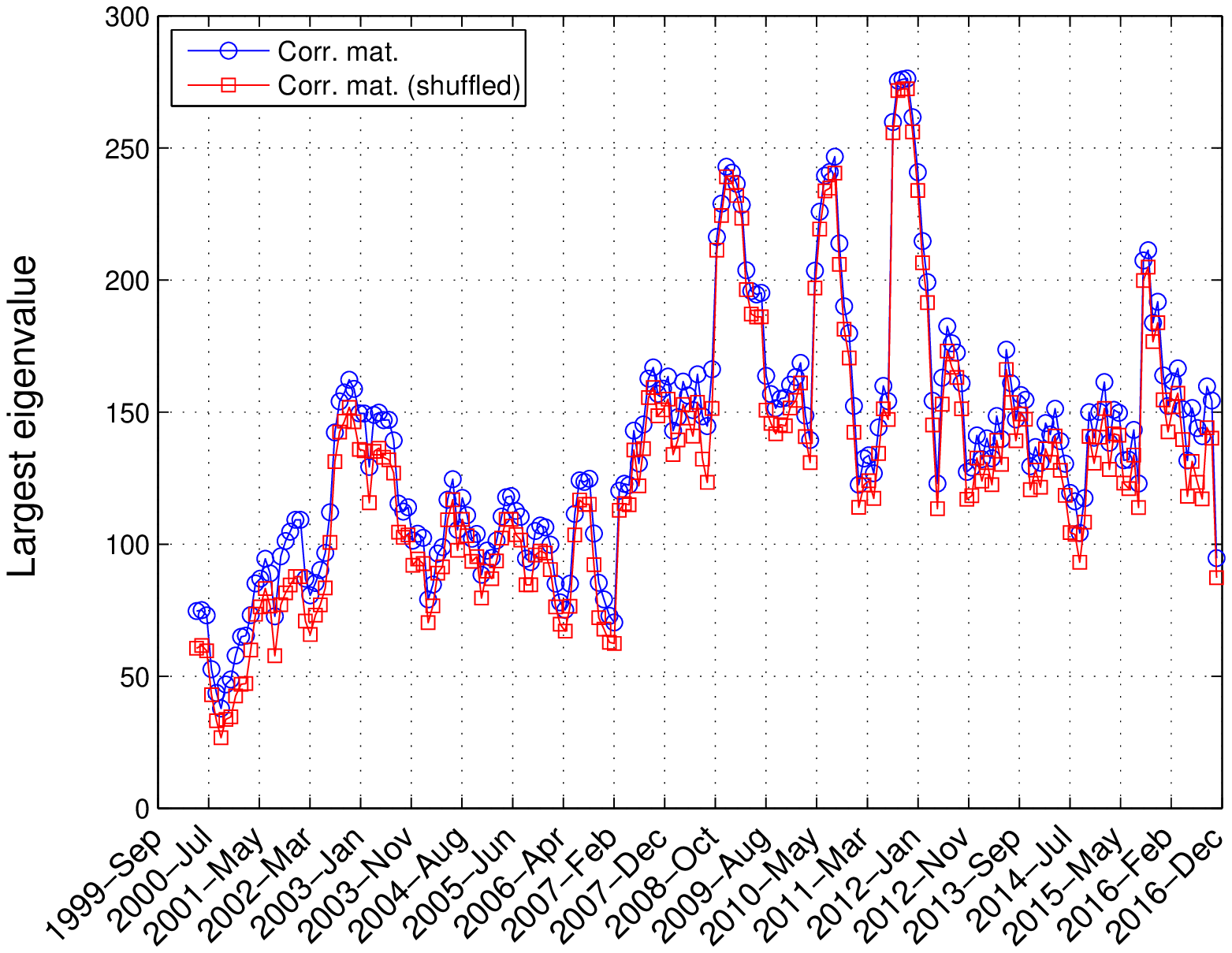}
\caption{The largest eigenvalues of the original correlation matrix (open circles) and its shuffled (open squares), for 90 working days window length. Irrespective of the obvious structural difference,  they both follow each other and signal the crises similarly.}
\label{fig:bigeigenvalue}
\end{figure}

During the financial crises, usually because of the spreading fear in the market, the stocks move together and correlation grows amongst them. This fact raises a natural question if our main result is just another derivation of the previous observations concerning the growth of
correlation between the indices. In order to address this appropriately, let us compute the largest eigenvalue of the correlation matrix \cite{lai, marsili, stanleyRMT}. As shown in Fig. \ref{fig:bigeigenvalue}, in accordance with our previous observations, over the crises the largest eigenvalue grows significantly. But this time, when we look at the largest eigenvalues of the "shuffled" correlation matrix, they behave exactly as in the original (unshuffled) matrix, i.e., in both cases the largest eigenvalues significantly grow over the crises (see Fig. \ref{fig:bigeigenvalue}). This means that the largest eigenvalues of the correlation matrix do not necessarily carry information about the structure of the interaction network. This is while in our analysis, we observe two totally different behaviors between the interaction network and
its shuffled one which provides a systematic way for a structural differentiation. Therefore we conclude that the off-critically over the crises is not a simple consequence of the growth in the correlations.

It should be notified that our observation close to the crises does not contradict the efficient market hypothesis, since it is not still clear if it can help one to extract money. Should we be able to extract money from such structure is left as an open question for the future works.

Extraction of real system's interaction networks is a rapid growing field. Beside the statics features of these networks, they could be used to deduce some dynamic features of the system. In this work, we established a correspondence between the critical phenomena and the external macroscopic state of a system. This approach could however be generalized to other fields where maximum entropy network is used like gene regulatory networks, neural networks, protein interactions and etc.

\section{Author contributions statement}

A.H.S., A.A.S., A.H. and E.A. designed the project. A.A.S. proposed the analysis and statistical methods. A.H.S. and P.T.S. has done the numerical computations. A.H.S., A.A.S. and A. H. analyzed the results and drafted the manuscript. 

\section{Additional information}

\subsection{Competing financial interests} 

The author(s) declare no competing financial interests.

\section{Data Availability}

The datasets analysed during the current study are available in the Yahoo Finance, http://finance.yahoo.com.


\end{document}